\documentclass{article}
\usepackage{spconf,amsmath,graphicx}
\usepackage{cite}

\usepackage{algorithm} 
\usepackage{algorithmic}  
\usepackage[algo2e]{algorithm2e}

\usepackage{graphicx}
\usepackage{amssymb}
\usepackage{indentfirst}
\usepackage{xspace}
\usepackage{amsmath}
\usepackage{blkarray, bigstrut}
\usepackage{tabstackengine}
\usepackage{xcolor,soul}
\usepackage{adjustbox}
\usepackage{bm} 
\usepackage{color} 
\usepackage{array}
\usepackage{balance}
\usepackage{multirow}
\usepackage{marginnote} 
\usepackage[norndcorners,customcolors,nofill]{hf-tikz}
\usepackage{amsmath,times,tabstackengine}
\usepackage[colorlinks=false,linkcolor=black,urlcolor=black,bookmarksopen=true, hidelinks]{hyperref}
\usepackage{booktabs}

\title{Time-Distributed feature learning in Network Traffic classification for Internet of Things}
\name{Yoga Suhas Kuruba Manjunath, Sihao Zhao, Xiao-Ping Zhang\thanks{This work was supported in part by the Natural Sciences and Engineering Research Council of Canada (NSERC), Grant No. RGPIN-2020-04661.}}
\address{Department of Electrical, Computer and Biomedical Engineering, Ryerson University, Canada\\
\{yoga.kuruba, sihao.zhao, xzhang\}@ryerson.ca}
\begin{document}
%
\maketitle
\begin{abstract}

The plethora of Internet of Things (IoT) devices leads to explosive network traffic. The network traffic classification (NTC) is an essential tool to explore behaviours of network flows, and NTC is required for Internet service providers (ISPs) to manage the performance of the IoT network. We propose a novel network data representation, treating the traffic data as a series of images. Thus, the network data is realized as a video stream to employ time-distributed (TD) feature learning. The intra-temporal information within the network statistical data is learned using convolutional neural networks (CNN) and long short-term memory (LSTM), and the inter pseudo-temporal feature among the flows is learned by TD multi-layer perceptron (MLP). We conduct experiments using a large data-set with more number of classes. The experimental result shows that the TD feature learning elevates the network classification performance by 10\%.

\end{abstract}
\section{Introduction}
\label{sec:intro}

The architecture of the interconnection of devices, which can send and receive the data, is termed as the Internet of Things (IoT) \cite{2015TowardsAD}. Emerging communication technologies, lightweight protocols, algorithms, and applications are allowing the organizations to design numerous IoT nodes. Norton has predicted around 21 billion IoT devices by 2025 \cite{Norton2019}. The fact hints a constant increase in the number of IoT devices.

The network traffic management (NTM) is a common technique used by internet service providers (ISP) to understand network behaviour and may be used to tune the performance of the network to reach the diverse class of users. The network traffic classification (NTC) is one of the essential components in the NTM. The NTC helps in categorizing the traffic flows, keeping account of the type of flow in the network. Hence the NTC plays a vital role in the Quality of Service (QoS), policy-shaping, and security operation \cite{Guan2000}. Besides, traffic classification helps in service level agreements (SLA) verification. Finally, ISPs must perform the lawful interception (LI) of illegal or critical traffic \cite{IETF3924}. Thus the ISPs should be aware of the type of content transmitted over their networks \cite{Valenti2013}. It is pivotal to classify the ever-increasing IoT traffic to understand the action of the IoT devices. The ISPs require NTC for performance tuning and policy-shaping.

In recent days, deep learning algorithms witnessed excellent results in network traffic classification \cite{8340067, 7932863}.  
Lopez-Martin \textit{et al.} \cite{8026581} uses the 
convolutional neural network (CNN), long short term memory (LSTM) and combination of CNN and LSTM to classify a private traffic data-set that has 6 features, 108 services, and 266,160 network flows, and achieves 96.32\% accuracy. Tong \textit{et al.} \cite{8647128} propose a CNN based classifier, which achieves 99.34\% of F1 for only 5 services with around 20,000 flows. Aceto \textit{et al.} \cite{8506558} use a 1D-CNN model to classify mobile encrypted traffic from Android, iOS, and Facebook, and obtains a 96.50\% accuracy. Yao \textit{et al.} \cite{8651277} find that the capsule neural network outperforms the CNN and CNN+LSTM for automatic feature selection to classify IoT traffic for smart cities. Their data-set contains 203,463 data flows, and 20 classes. Lotfollahi \textit{et al.} \cite{deeppacket} employ the sparse autoencoder (SAE) and CNN to classify an encrypted data-set. They use 1,500 bytes from each Maximum Transmission Unit (MTU) to train the deep packet model that produces 98\% of accuracy.

We observe the researches employ CNN-LSTM based model \cite{8647128,8651277,a1,a2,a3}. However, the data used in the studies are not available; therefore we implemented the proposed models and tested using our data and it found that model performance was worse on our data. Those existing researches have used the limited number of traffic and few classes. Comparison of the classification performance is difficult since the datasets used are different. In this work, our objective is to investigate the effectiveness of time-distributed feature learning in comparison with the vanilla deep learning models. The IoT traffic is random and diverse \cite{7968561}. Therefore, it is possible that the models would not scale and might fail in classifying diverse IoT traffic.  

In this paper, we propose i) a novel representation of network traffic data to enable time-distributed (TD) feature learning, and ii) a deep learning model with TD feature learning for better classification of the network traffic.

We use the combination of CNN and LSTM to extract the temporal information within the flows. The output of CNN and LSTM is modelled as a video stream to apply TD feature learning using TD layer over MLP to maximize the performance of deep learning model. With our experiment, it is shown that TD layer over MLP helps to find the pseudo-temporal features that is not identified by the LSTM.

We prove the performance of TD feature learning by comparing the model with and without TD layer. It is observed that the model with TD feature learning achieves over 95\% accuracy with novel data representation. On the other hand, the model without the TD feature learning achieves an accuracy over 80\%. We use a real network traffic that consists of millions of flow and includes diverse application traffic. The proposed solution classifies more than 135 application traffic. Based on our literature work, this is the first work to employ TD feature learning for network traffic classifier.

\section{Novel representation of network flows}
\label{wd}

\begin{figure}
 	\centering
 	\includegraphics[width=0.99\linewidth]{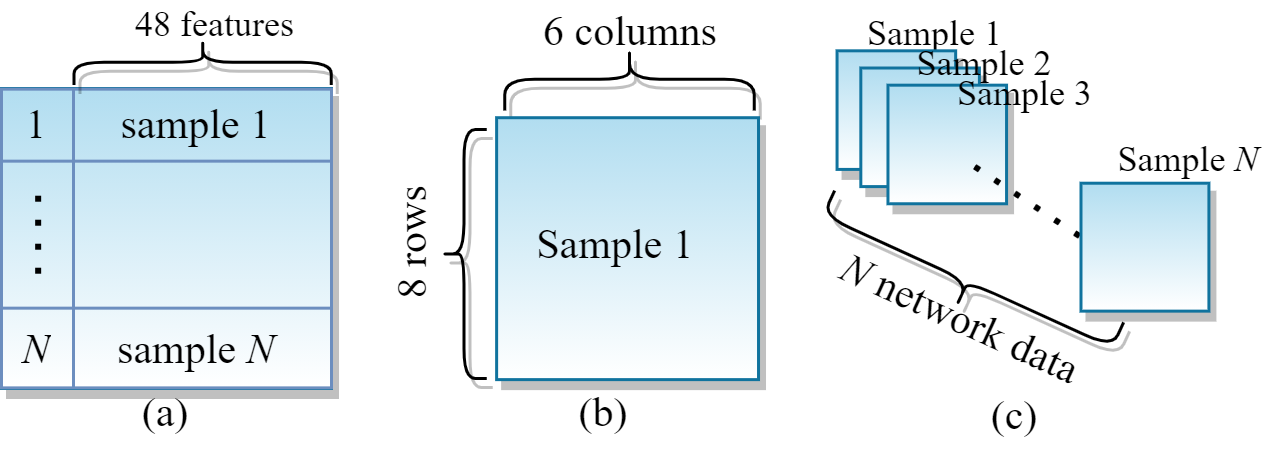} 
 	\caption{(a) Data in the unprocessed format, $N$ is the total number of samples. (b) Representation of each sample as a 2D grey-scale image. (c) Data as a series of grey-scale images}
 	\label{fig:data}
\end{figure}

We use the IoT traffic data collected in the network section of Universidad Del Cauca, Popayán, Colombia \cite{kaggle}. Around 2,704,839 numbers of flows that are highly unbalanced from 141 different classes collected at various hours are used in the experiment. The original data-set consists of 48 different features the represents the statistical information of flows such as source/destination IP, source/destination port, minimum, average, and maximum of inter-arrival time of given flow, and many. The associated network service extracted using deep packet inspection (DPI) is used as the label. The data in unprocessed format is exhibited in Fig. \ref{fig:data} (a). 

The traffic flows consist of neighbourhood information from the definition of protocols from the open systems interconnection (OSI) layers. The pattern between the successive data is captured by nature. Note that if the flow is represented in the image form, it is possible to extract the extra inter layer information, which can possibly lead to a better classification.


We represent each sample of data to a 2D matrix as shown in Fig. \ref{fig:data} (b). We denote the original $i$-th sample as shown in Fig. \ref{fig:data} (a) by vector $\boldsymbol{x}_i^T$, and the $i$-th sample after processing by matrix $\bm{Y}_i$. Their relation is $\left[\bm{Y}_i\right]_{j,:}=\left[\boldsymbol{x}_i\right]_{(j-1)W+1:jW}^T$, $i=1,\cdots,K$, where $K$ and $W$ are the factors of the length of $\boldsymbol{x}_i$, $[\cdot]_{j,:}$ is the $j$-th row of a matrix and $[\cdot]_{m:n}$ is a vector formed by the $m$-th to the $n$-th elements of a vector. Algorithm \ref{Al:algorithm1} shows the steps to represent the given network flows into a video stream.



\begin{algorithm}
\caption{Algorithm of the novel 2D representation of network flows}\label{Al:algorithm1}

\KwData{Input: $\boldsymbol{x}_i^T$}
\KwResult{Output: $\bm{Y}_i$}
 initialization\;
 
    Choose $K$ and $W$ that are factors of  the length of $\boldsymbol{x}_i$
  
 \While{i $\leqslant$ K}{
    $\left[\bm{Y}_i\right]_{j,:}=\left[\boldsymbol{x}_i\right]_{(j-1)W+1:jW}^T$
  }

\end{algorithm}

With the above-mentioned novel representation, we can employ the TD feature learning \cite{8026581}. In the experiment each flow consists of 48 features; therefore, we choose $K = 8$ and $W = 6$, which is the nearest representation of a square image as shown in Fig. \ref{fig:data} (b). Hence, we realise all flows as a series of images that is treated as a video stream. The representation of network flows as a series of grey-scale images is shown in Fig. \ref{fig:data} (c). According to \cite{10.1145/1282380.1282400}, the proposed network data representation is vital to extract the weak correlation and pseudo-temporal information that is present in the network traffic.

\section{Model with TD feature Learning}

In this section, we propose a supervised deep learning model that consists of CNN, LSTM and MLP that uses TD feature learning and is termed as Conv-LSTM-TD(MLP).
We present a deeper model that uses four layers of CNN and a LSTM layer to handle big data. CNN requires its input data in an image format, and therefore we represent the network data into an image as presented in the previous section. CNN extracts the local neighbourhood information that is important to classify a network flow. The feature maps extracted from CNN are fed to LSTM. LSTM finds a temporal information present in the feature maps extracted by CNN. The LSTM provides a sequence of information, which is seen with an additional temporal axis that represents a video stream. This enables us to apply the TD feature learning to extract temporal information that is present across the sequence.

\begin{figure} 
 	\centering
 	\includegraphics[width=0.99\linewidth]{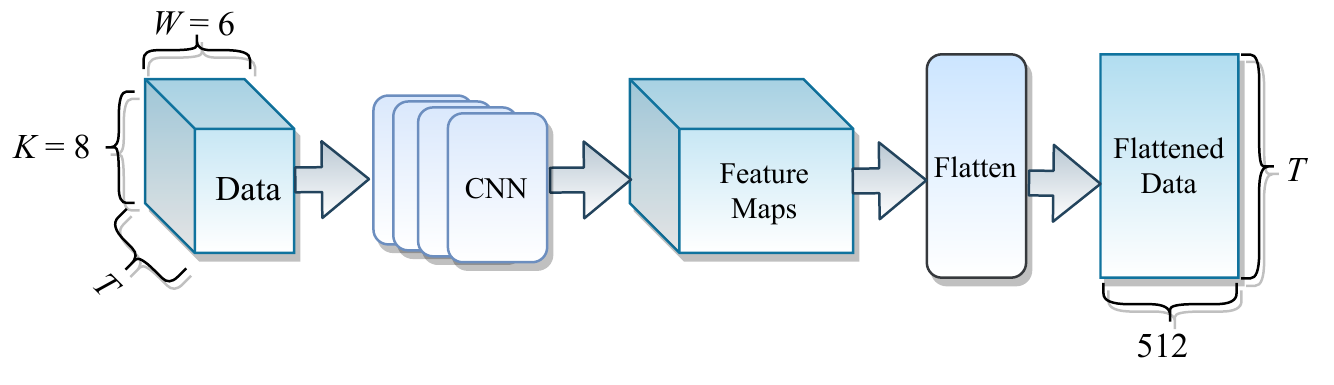}
 	\caption{CNN model. $T$ is the number of the training samples. Neighbourhood feature is extracted by using 4 layers of CNN with 2D maxpool and batch normalization. The multi-dimensional tensor is flattened to prepare the input to LSTM}
 	\label{fig:CNN model}
\end{figure}

 The CNN model used in the proposed model is shown in Fig. \ref{fig:CNN model}. Each layer of CNN has 64 filter with $(2\times2)$ kernels. A 2D maxpool and batch normalization are used in one CNN layer to regularize the model to avoid over-fitting. All layers of CNN uses rectified linear unit (ReLU) activation function to avoid the saturation issue during back propagation \cite{xu2015empirical, Kent2019PerformanceOT}. 

\begin{figure}
 	\centering
 	\includegraphics[width=0.99\linewidth]{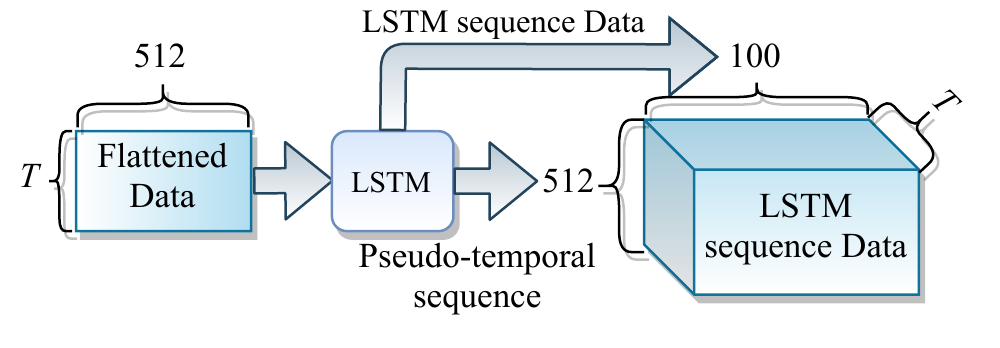}
 	\vspace{-0.2cm}
 	\caption{LSTM model is used to extract temporal feature from CNN's feature maps. The pseudo-temporal sequence is added by extracting the hidden layer's output of LSTM}
 	\label{fig:LSTM model}
\end{figure}

The LSTM is employed to find the time correlated information from the feature maps provided by the CNN. We use 100 units of LSTM with ReLU as the activation function. We extract the full sequence, namely pseudo-temporal sequence, from the hidden units of LSTM, which acts as a temporal axis. This is achieved by using the return sequence in Tensorflow. The LSTM output with the pseudo-temporal sequence is seen as a video stream, as shown in Fig. \ref{fig:LSTM model}.

\begin{figure}
 	\centering
 	\includegraphics[width=0.99\linewidth]{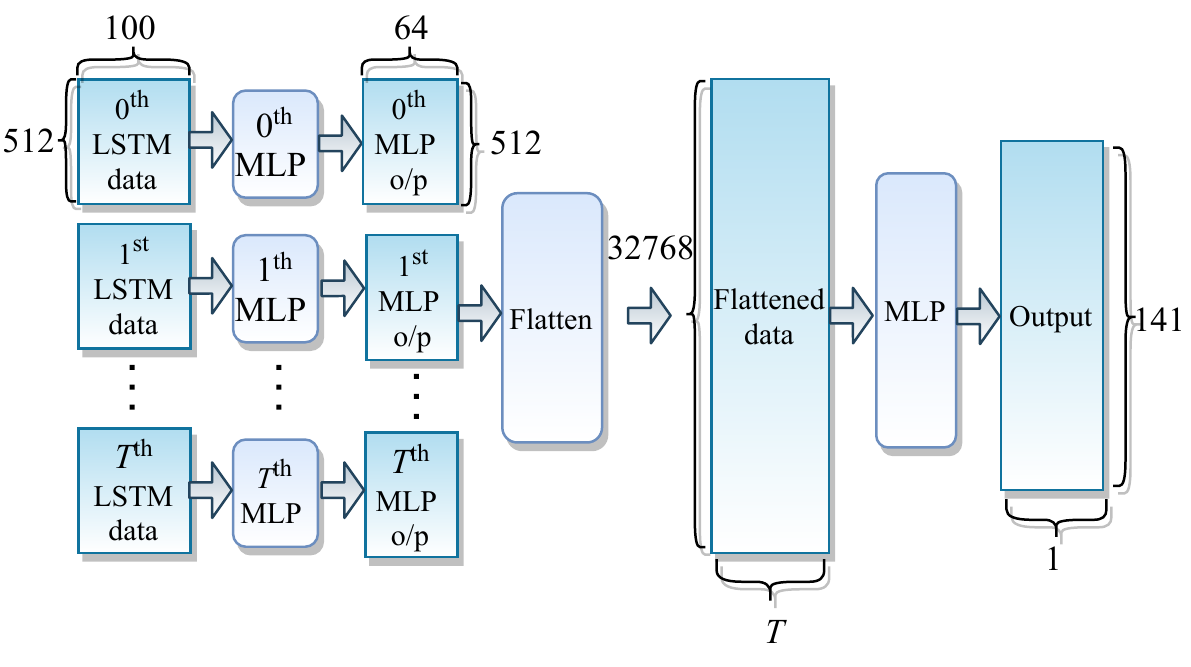} 
 	\caption{Proposed TD MLP model. The TD wrapper applies each layer of MLP to time sequence from the LSTM sequence data. Thus, the TD feature learning is employed to learn the pseudo-temporal features.}
 	\label{fig:Timedistributed MLP model}
\end{figure}

The TD wrapper is applied to the MLP layer to enable TD feature learning. The applied TD wrapper applies the MLP to every pseudo-temporal sequence of output from the LSTM as shown in Fig. \ref{fig:Timedistributed MLP model}. Thus, it leads to more training time as well as more parameters to train. However, this helps in improving the classification performance \cite{qiao2018time}, and will be shown in Section \ref{chap:Results}. 

The MLP with TD wrapper consists of 64 units and ReLU as the activation function. The output of TD MLP is flattened to generate input to the last MLP with the number of units equal to 141, i.e., the number of classes. The last layer of MLP uses softmax as the activation to derive the output as a probability distribution. Therefore, we use the cross-entropy loss function \cite{Rumelhart1986LearningRB} and the Adam optimizer with an adaptive learning rate \cite{186212}. The model hyper-parameters tuning is done based on the model performance on validation data-set.

\section{parameter analysis of TD model}
\label{sec:4}

The structure of the proposed new Conv-LSTM-TD(MLP) model and the number of parameters in each layers are listed in Table \ref{table:Calculation of parameters of Conv-LSTM-TD(MLP) model.}. The number of learn-able parameters of the CNN is given by $ (V \times H\times S + B) \times U$, where $V$ and $H$ are the vertical and horizontal size of the kernel, respectively, $S$ is the input size and $U$ is the number of units and $B$ is bias. Batch normalization (BN) consists of gamma and beta weights, which are trainable, and moving mean as well as moving variance, which are non-trainable parameters \cite{10.5555/3045118.3045167}. The number of trainable parameters of BN is $ 2 \times U$. For LSTM, the number of trainable parameters is $4\times\left[(S+1) \times U + U^2 \right]$. LSTM includes three sigmoids and one hyperbolic tangent functions \cite{Goodfellow-et-al-2016} hence the first term is 4 in the above expression. In an MLP, $S \times U + U$ gives the number of parameters to train.


The last MLP layer in proposed model receives 512$\times$64 additional parameters because of TD feature learning. Therefore, in case of TD feature learning, if $L$ number of CNNs are used in the model with the last layer of CNN having $U_L$ units, then $V_{L}\times H_{L}\times U_{L}$ gives an additional pseudo-temporal sequence, where $V_L$ and $H_L$ are the vertical and horizontal size of the kernel of the last CNN with $U_L$ units. The LSTM extracts the temporal sequence that is equal to 512 given in $V_{L}\times H_{L}\times U_{L}$. We take a sum of the last column of the table and find that the Conv-LSTM-TD(MLP) has 4,726,189 trainable parameters in total. In the Conv-LSTM-MLP, model without TD feature learning, the last MLP layer consists of $64\times 141 + 141$ number of parameters. Thus, the Conv-LSTM-MLP has 114,925 train-able parameters. We observe that Conv-LSTM-TD(MLP) model has nearly 41 times more parameters to train in comparison with Conv-LSTM-MLP model.


\begin{table}
\caption{Structure of Conv-LSTM-TD(MLP) and number of trainable parameters.}
\centering 
\begin{tabular}{
>{\centering\arraybackslash}m{1.5cm} 
>{\centering\arraybackslash}m{3.9cm} >{\centering\arraybackslash}m{1.6cm}} 
\hline
Network & Calculation & Trainable parameters \\
\hline 
CNN\_2D\_0 & $(2\times2\times1+1)\times64$ & 320\\
MP\_2D\_0 & - & 0 \\
BN\_0 & $2\times64$ &  128\\
CNN\_2D\_1 & $(2\times2\times64+1)\times64$& 16448 \\
MP\_2D\_1 &  -& 0 \\
BN\_1 & $2\times64$ &  128 \\
CNN\_2D\_2 &$(2\times2\times64+1)\times64$& 16448\\
MP\_2D\_2 &-& 0\\
BN\_2 & $2\times64$ &  128 \\
CNN\_2D\_3 &$(2\times2\times64+1)\times64$& 16448 \\
MP\_2D\_3 &-& 0\\
BN\_3 &$2\times64$ &  128 \\
Flatten & - & 0 \\
LSTM & $4\times\left[(1+1)\times100+100^2\right]$ & 40800\\
TD(MLP\_0) & $100\times64+64$ & 6464\\
TD(MLP\_1) & $64\times64+64$ & 4160 \\
TD(MLP\_2) & $64\times64+64$ & 4160 \\
Flatten & - & 0 \\
MLP\_3 & $512\times64\times141+141$ & 4620429 \\

\hline
\end{tabular}
\label{table:Calculation of parameters of Conv-LSTM-TD(MLP) model.}
\end{table}

\section{Experimental Results}
\label{chap:Results}

In this section, we investigate the performance of the TD feature learning on the IoT traffic flows in-terms of training time as well as accuracy. 

We use Tensorflow library to implement the software and Nvidia GeForce RTX 2080 Super with Max-Q graphical processing unit to train the model. We use 80\% of the data, i.e., 2,163,871 flows, to train the model, and 10\% of the data, i.e., 270,483 flows of the data for validation. The rest 10\%, i.e., 270,485 flows of the data is used to test the model. In addition, 10\% of the validation set is used to choose the model hyper-parameters. We employ early stopping to stop training when validation error starts to increase. Thus, we obtain better model.

The training process in the experiment shows that the Conv-LSTM-TD(MLP) model takes an average of 520 seconds for each epoch and around an average of 45 number of epochs. On the other hand, the Conv-LSTM-MLP model takes an average of 240 seconds for each epoch and an average of 56 number of epochs. The proposed Conv-LSTM-TD(MLP) model takes nearly twice the training time of Conv-LSTM-MLP model. Table \ref{table:Training time comparision of Conv-LSTM-MLP and Conv-LSTM-TD(MLP) models.} shows the training time comparison of the model with and without time distributed feature learning.

The effectiveness of TD feature learning is investigated using two models with and without TD learning. The two models are Conv-LSTM-TD(MLP) and Conv-LSTM-MLP. The classification performance is evaluated using the following 4 performance metrics for each model.

\begin{table}
\caption{Training time comparison of Conv-LSTM-MLP and Conv-LSTM-TD(MLP) models. TD wrapper needs twice the training time due to more parameters to train.}
\centering 
\begin{tabular}{
>{\centering\arraybackslash}m{3.1cm} 
>{\centering\arraybackslash}m{1.7cm} >{\centering\arraybackslash}m{2cm}} 
\hline
Model & Avg. time per epoch & Avg. number of epochs \\
\hline 
Conv-LSTM-TD(MLP) & 520 sec & 45\\
Conv-LSTM-MLP & 240 sec & 56\\

\hline
\end{tabular}
\label{table:Training time comparision of Conv-LSTM-MLP and Conv-LSTM-TD(MLP) models.}
\end{table}

\begin{align}
\label{Equ:performance metrics}
    Accuracy &= \frac{TP+TN}{TP+TN+FP+FN}\nonumber\\
    Precision &= \frac{TP}{TP+FP}\nonumber\\
    Recall &= \frac{TP}{TP+FN}\nonumber\\
    F1 &= 2\times \frac{Precision\times Recall}{Precision+Recall} \nonumber
\end{align}
where false-positive (FP) represents a false detection when there is actually no detection. False-negative (FN) indicates no detection when there is one. True-positive (TP) depicts a correct detection. True-negative (TN) represents no detection correctly. F1 is the harmonic mean of precision and recall. Therefore it gives information about the correct detection. Higher F1 score indicates better performance.









\begin{figure}
	\centering
	\includegraphics[width=0.99\linewidth]{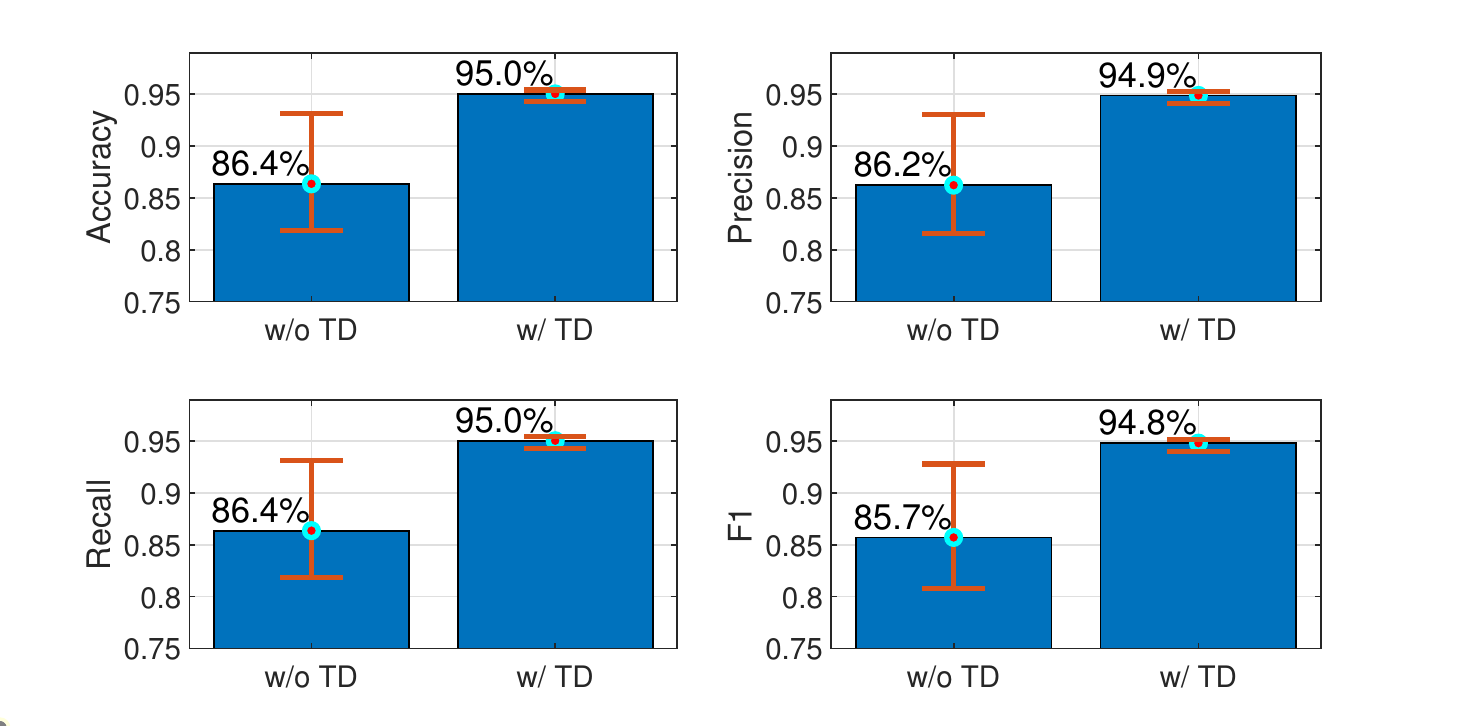}
	\caption{Performance metrics of the proposed models with and without the TD wrapper. The average values with error at each bar are from the 5 experiments. The performance of the proposed Conv-LSTM-TD(MLP) with the TD wrapper is about 10\% better than that of the one without.}
	\label{fig:results}
\end{figure}

We train and test both the models for 5 times, to obtain an overall performance. The average values of the performance metrics on test data as well as their error ranges for the two models from 5 times of experiments are shown in Fig. \ref{fig:results}. We can see that the performances for the 5-time tests varies a bit. However, the average values clearly show that the model with the TD wrapper, i.e., Conv-LSTM-TD(MLP), outperforms the model without, i.e., Conv-LSTM-MLP, consistent with the analysis presented in the previous sections. It also demonstrates that the proposed model with CNN+LSTM along with MLP employed with TD feature learning is effective in learning the inter and intra-temporal features among the network traffic flows and provides 95\% NTC accuracy. Table \ref{table:Training time comparision of Conv-LSTM-MLP and Conv-LSTM-TD(MLP) models.} shows that training time for TD based model is almost twice than training the vanilla model because of more parameters as explained in Section \ref{sec:4}. However, the performance of TD based model is observed to be 10\% better than vanilla model as shown in Fig. \ref{fig:results}; therefore training time can be considered to trade off for a better performance.

\section{Conclusion and future work}
In the era of IoT, NTC plays a vital role for internet service providers to tune the network performance. We contribute to the NTC by proposing a novel data representation for network flows and a deep learning model with TD feature learning, namely Conv-LSTM-TD(MLP). It employs CNN, LSTM, and the MLP with TD wrapper to learn the intra and inter-flow features of the network data. We expect that the proposed solution helps the ISPs to tune the service based on the traffic type, and thus, improve the quality of service.

To the best of our knowledge, this is the first work to represent the network data as video stream, and employ the TD feature learning to extract pseudo-temporal correlation for a better classification of IoT traffic flows. Through the experiment, we show that TD feature learning elevates the NTC performance by 10\%. The proposed model with TD feature learning achieves 95\% accuracy. Thus, we find that the novel representation of traffic flows and the TD feature learning revamp the NTC for IoT.

We consider to study the impact of parameters to investigate the scale of the model. The optimal convergence rate of the solution in the case of both on and off line deployment needs to be studied to achieve practical fruition of the proposed solution.


\bibliographystyle{IEEEtran}
\balance
\bibliography{IEEEabrv,refs}

\begin{thebibliography}{10}
\providecommand{\url}[1]{#1}
\csname url@samestyle\endcsname
\providecommand{\newblock}{\relax}
\providecommand{\bibinfo}[2]{#2}
\providecommand{\BIBentrySTDinterwordspacing}{\spaceskip=0pt\relax}
\providecommand{\BIBentryALTinterwordstretchfactor}{4}
\providecommand{\BIBentryALTinterwordspacing}{\spaceskip=\fontdimen2\font plus
\BIBentryALTinterwordstretchfactor\fontdimen3\font minus
  \fontdimen4\font\relax}
\providecommand{\BIBforeignlanguage}[2]{{%
\expandafter\ifx\csname l@#1\endcsname\relax
\typeout{** WARNING: IEEEtran.bst: No hyphenation pattern has been}%
\typeout{** loaded for the language `#1'. Using the pattern for}%
\typeout{** the default language instead.}%
\else
\language=\csname l@#1\endcsname
\fi
#2}}
\providecommand{\BIBdecl}{\relax}
\BIBdecl

\bibitem{2015TowardsAD}
R.~Minerva, A.~Biru, and D.~Rotondi, ``Towards a definition of the {Internet of
  Things (IoT)},'' \emph{IEEE Internet Initiative}, vol.~1, no.~1, pp. 1--86,
  2015.

\bibitem{Norton2019}
\BIBentryALTinterwordspacing
S.~Symanovich, ``The future of {IoT}: 10 predictions about the {Internet of
  Things}.'' [Online]. Available:
  \url{https://us.norton.com/internetsecurity-iot-5-predictions-for-the-future-of-iot.html}
\BIBentrySTDinterwordspacing

\bibitem{Guan2000}
Q.~Guan, ``Quality of services for {ISP} networks,'' in \emph{International
  Conference on Intelligence in Networks}.\hskip 1em plus 0.5em minus
  0.4em\relax Springer, 2000, pp. 471--483.

\bibitem{IETF3924}
F.~Baker, B.~Foster, and C.~Sharp, ``The {Cisco} architecture for lawful
  intercept in {IP} networks,'' Available at
  \url{https://tools.ietf.org/html/rfc3924}.

\bibitem{Valenti2013}
S.~Valenti, D.~Rossi, A.~Dainotti, A.~Pescap{\`e}, A.~Finamore, and M.~Mellia,
  ``Reviewing traffic classification,'' in \emph{Data Traffic Monitoring and
  Analysis}.\hskip 1em plus 0.5em minus 0.4em\relax Springer, 2013, pp.
  123--147.

\bibitem{8340067}
L.~{Liu}, B.~{Yin}, S.~{Zhang}, X.~{Cao}, and Y.~{Cheng}, ``Deep learning meets
  wireless network optimization: Identify critical links,'' \emph{{IEEE} Trans.
  Netw. Sci. Eng.}, vol.~7, no.~1, pp. 167--180, 2020.

\bibitem{7932863}
Z.~M. {Fadlullah}, F.~{Tang}, B.~{Mao}, N.~{Kato}, O.~{Akashi}, T.~{Inoue}, and
  K.~{Mizutani}, ``State-of-the-art deep learning: Evolving machine
  intelligence toward tomorrow’s intelligent network traffic control
  systems,'' \emph{{IEEE} Commun. Surveys Tuts.}, vol.~19, no.~4, pp.
  2432--2455, 2017.

\bibitem{8026581}
M.~Lopez-Martin, B.~Carro, A.~Sanchez-Esguevillas, and J.~Lloret, ``Network
  traffic classifier with convolutional and recurrent neural networks for
  {Internet of Things},'' \emph{IEEE Access}, vol.~5, pp. 18\,042--18\,050,
  2017.

\bibitem{8647128}
V.~{Tong}, H.~A. {Tran}, S.~{Souihi}, and A.~{Mellouk}, ``A novel quic traffic
  classifier based on convolutional neural networks,'' in \emph{2018 IEEE
  Global Communications Conference (GLOBECOM)}, 2018, pp. 1--6.

\bibitem{8506558}
G.~{Aceto}, D.~{Ciuonzo}, A.~{Montieri}, and A.~{Pescapé}, ``Mobile encrypted
  traffic classification using deep learning,'' in \emph{2018 Network Traffic
  Measurement and Analysis Conference (TMA)}, 2018, pp. 1--8.

\bibitem{8651277}
H.~{Yao}, P.~{Gao}, J.~{Wang}, P.~{Zhang}, C.~{Jiang}, and Z.~{Han}, ``Capsule
  network assisted {IoT} traffic classification mechanism for smart cities,''
  \emph{{IEEE} Internet Things J.}, vol.~6, no.~5, pp. 7515--7525, 2019.

\bibitem{deeppacket}
M.~{Lotfollahi}, M.~{Jafari Siavoshani}, and R.~{Shirali Hossein Zade}, ``Deep
  packet: a novel approach for encrypted traffic classification using deep
  learning,'' \emph{Soft Computing}, vol.~24, 2020.

\bibitem{a1}
H.~{Lim}, J.~{Kim}, J.~{Heo}, K.~{Kim}, Y.~{Hong}, and Y.~{Han}, ``Packet-based
  network traffic classification using deep learning,'' in \emph{2019
  International Conference on Artificial Intelligence in Information and
  Communication (ICAIIC)}, 2019, pp. 046--051.

\bibitem{a2}
S.~{Rezaei} and X.~{Liu}, ``Deep learning for encrypted traffic classification:
  An overview,'' \emph{IEEE Communications Magazine}, vol.~57, no.~5, pp.
  76--81, 2019.

\bibitem{a3}
G.~{Aceto}, D.~{Ciuonzo}, A.~{Montieri}, and A.~{Pescapé}, ``Mobile encrypted
  traffic classification using deep learning: Experimental evaluation, lessons
  learned, and challenges,'' \emph{IEEE Transactions on Network and Service
  Management}, vol.~16, no.~2, pp. 445--458, 2019.

\bibitem{7968561}
I.~{Florea}, R.~{Rughinis}, L.~{Ruse}, and D.~{Dragomir}, ``Survey of
  standardized protocols for the internet of things,'' in \emph{2017 21st
  International Conference on Control Systems and Computer Science (CSCS)},
  2017, pp. 190--196.

\bibitem{kaggle}
\BIBentryALTinterwordspacing
J.~S. Rojas, ``Labeled network traffic flows - 141 applications,'' accessed =
  2020-04-07. [Online]. Available:
  \url{https://www.kaggle.com/jsrojas/labeled-network-traffic-flows-114-applications}
\BIBentrySTDinterwordspacing

\bibitem{10.1145/1282380.1282400}
S.~Chachulski, M.~Jennings, S.~Katti, and D.~Katabi, ``Trading structure for
  randomness in wireless opportunistic routing,'' \emph{ACM SIGCOMM Computer
  Communication Review}, vol.~37, no.~4, pp. 169--180, 2007.

\bibitem{xu2015empirical}
B.~Xu, N.~Wang, T.~Chen, and M.~Li, ``Empirical evaluation of rectified
  activations in convolutional network,'' \emph{arXiv preprint
  arXiv:1505.00853}, 2015.

\bibitem{Kent2019PerformanceOT}
D.~Kent and F.~M. Salem, ``Performance of three slim variants of the long
  short-term memory ({LSTM}) layer,'' \emph{2019 IEEE 62nd International
  Midwest Symposium on Circuits and Systems (MWSCAS)}, pp. 307--310, 2019.

\bibitem{qiao2018time}
H.~Qiao, T.~Wang, P.~Wang, S.~Qiao, and L.~Zhang, ``A time-distributed
  spatiotemporal feature learning method for machine health monitoring with
  multi-sensor time series,'' \emph{Sensors}, vol.~18, no.~9, p. 2932, 2018.

\bibitem{Rumelhart1986LearningRB}
D.~E. Rumelhart, G.~E. Hinton, and R.~J. Williams, ``Learning representations
  by back-propagating errors,'' \emph{Nature}, vol. 323, pp. 533--536, 1986.

\bibitem{186212}
T.~Chilimbi, Y.~Suzue, J.~Apacible, and K.~Kalyanaraman, ``Project adam:
  Building an efficient and scalable deep learning training system,'' in
  \emph{11th USENIX Symposium on Operating Systems Design and Implementation
  (OSDI' 14)}, 2014, pp. 571--582.

\bibitem{10.5555/3045118.3045167}
S.~Ioffe and C.~Szegedy, ``Batch normalization: Accelerating deep network
  training by reducing internal covariate shift,'' in \emph{Proceedings of the
  32nd International Conference on International Conference on Machine Learning
  - Volume 37}, ser. ICML'15.\hskip 1em plus 0.5em minus 0.4em\relax JMLR.org,
  2015, p. 448–456.

\bibitem{Goodfellow-et-al-2016}
A.~Gulli and S.~Pal, \emph{Deep learning with Keras}.\hskip 1em plus 0.5em
  minus 0.4em\relax Packt Publishing Ltd, 2017.

\end{thebibliography}

\end{document}